# Philosophical aspects of modern cosmology[1]

Henrik Zinkernagel
Department of Philosophy I, Campus de Cartuja,
18071, University of Granada, Spain.
zink@ugr.es

Cosmology is the attempt to understand in scientific terms the structure and evolution of the universe as a whole. This ambition has been with us since the ancient Greeks, even if the developments in modern cosmology have provided a picture of the universe dramatically different from that of Pythagoras, Plato and Aristotle. The cosmological thinking of these figures, e.g. the belief in uniform circular motion of the heavens, was closely related to their philosophical ideas, and it shaped the field of cosmology at least up to the times of Copernicus and Kepler.

Nowadays it is not uncommon among scientists to question the relevance of philosophy for their field. This may be part of a simplified view according to which science is mostly about finding the best match between theories and empirical data. However, even on such a view one can identify interesting philosophical issues, like underdetermination of theories and theoryladenness of data. Moreover, apart from matching theory and data, science is often concerned with what the studied theories implies for our deeper understanding of the world. This involves the philosophical activity of interpreting the theories in question, and philosophy thus continues to be an integral part of scientific, including cosmological, thought. One may argue that cosmology is even more philosophical than most other sciences, in that it more explicitly deals with the limits or horizons of scientific knowledge. In particular, as cosmology involves the age-old questions of the possible temporal and spatial limits of the universe, it is naturally associated with irresistible speculations of what may cause or lie beyond those limits.

In the fall of 2011, the University of Granada hosted an interdisciplinary workshop – *Philosophical Aspects of Modern Cosmology* – which aimed to present and discuss new developments concerning methodological, conceptual, epistemological and metaphysical issues in modern cosmology. A basic characteristic of the workshop was the stimulating interchanges between historians and philosophers of science and observational and theoretical cosmologists. The papers in the present volume are the written outcome of this event.

Broadly speaking, the philosophy of cosmology is concerned with the foundational and philosophical assumptions and consequences of cosmological theories and observations. This includes philosophical issues surrounding the 'ingredient theories' of cosmology; first of all general relativity, but also cosmological implications or uses of quantum mechanics, quantum field theory and thermodynamics. Systematic surveys of the

---

philosophical aspects of modern cosmology are practically inexistent. The few possible exceptions to this include the overview articles by Ellis (2007), Smeenk (2013), and Ellis (this volume). The recent dates of these overviews bear witness to a currently rising interest in the philosophy of cosmology, and the fact that they are so few in number suggests that there is still much ground to be covered.

The papers in the present volume of course reflect the particular and different interests of the authors, and they cannot be taken to constitute a systematic survey of the field. It is to be hoped, nevertheless, that these papers may serve as inspiration for further exploration and mapping of the philosophical territory of cosmology. In the following, I provide a brief outline of the contents of the papers before closing with a few remarks on the current interest in philosophy of cosmology.

The volume opens with George Ellis' overview of some fundamental issues in philosophy of cosmology. These include the demand to, and scope of, explanations in cosmology, the implications of the limits of our observational access to distant domains of the universe, and the possible tests of alternative models and of consistency within the standard cosmological model. The paper provides a spirited defense of the relevance of philosophy for cosmological thought by one of cosmology's main practitioners. This is brought out clearly e.g. in Ellis' discussion of the uniqueness of the universe and of the multiverse – an attempt to deny this uniqueness. Regarding the multiverse, Ellis argues that its advocates unjustifiably invoke physically existing infinities and undermine the most important criterion for what constitutes genuine science – namely observational and experimental support (cf. also Kragh's contribution). A particular concern for Ellis is the scope of cosmology as a science. As he points out, one may in physical cosmology restrict attention to more technical issues regarding e.g. fits between theory and observation or modes of argumentation. On the other hand, if one expects cosmology to have a deeper impact on the way we see ourselves in the world – a sentiment often reflected in the titles, or at least on the back-flaps, of popular cosmology books – then one has to go into a very different territory where much more than physics is involved. Ellis argues that in this latter case one must take on the relation between cosmology and such big and open-ended issues as emergence of complexity, meaning, purposeful action and ultimate causation.

Dominico Giulini's paper is also concerned with the scope of cosmology, though in a quite different sense from the way this is addressed by Ellis and Kragh. While the multiverse discussion is often about a possible upper limit – or "outward bound" – for when cosmology is a science, Giulini addresses a possible lower limit – or "inward bound" – for when and for what objects cosmological (expansion) effects are relevant. Giulini's attack on this question is to investigate how the global expansion affects local physics. He first argues that the often invoked idea of expansion as a stretching of space is misleading. Rather, cosmological expansion is best thought of in terms of changes in the inertial structure of space, that is, as geodesics deviating from one another. Giulini shows how such a notion of expansion can be captured in a simple pseudo-Newtonian picture – as an extra term in Newton's second law (akin to what is done when rewriting this law for non-inertial reference frames). These considerations lead to the conclusion that gravitationally bound systems roughly below the scale of galaxy clusters (20-30 Mpc) do not expand as their gravitational binding force is larger than the 'expansion force'. The intuitive form of this Newtonian picture is later justified in the context of general relativity. Giulini finally takes up the difficult subject of how to describe the possible effects of expansion on objects such as black holes which require a full general relativistic treatment. Expansion may change the mass and horizon structure of a black hole but the issue is complicated since there is no natural way to superpose solutions (or interpolate spacetimes) in general relativity.

A conceptual discussion of the question of time in relativity and cosmology is presented in the paper by Marc Lachièze-Rey. He argues that no coherent notion of time – which unites the main characteristics usually attributed to time – can be found once we move to Einsteinian relativistic physics. While this claim is not novel, Lachièze-Rey illustrates his case with an interesting analogy between the change from Aristotelian to Newtonian physics (in which the vertical is given up) and the change between Newtonian and Einsteinian physics (in which time is in some sense given up). The discussion of the different aspects of time in relativity builds up to a discussion of the notion of cosmic time. Contrary to standard presentations of this topic, Lachièze-Rey rejects the usual way of introducing cosmic time via solutions to the Einstein equations which satisfy the cosmological principle, that is, the assumption of homogeneity and isotropy of the spatial sections. (For a different problem with this procedure, see Rugh and Zinkernagel 2011). Lachièze-Rey instead discusses a definition according to which cosmic time at an event is the supremum of all the proper durations of all future directed time-like curves ending in this event. While Lachièze-Rey points out that such a notion is not always well-defined (that is, finite), he argues that this definition is preferable when dealing with the real cosmological situation in which the universe is, of course, not perfectly homogeneous and isotropic. In any case, cosmic time is only one among several relevant, but mutually contradictory, choices for a time-function in cosmology – none of which, argues Lachièze-Rey, can be identified with time as such.

The following two papers address general epistemic issues in modern cosmology. Helge Kragh's contribution is on a possible epistemic shift in contemporary cosmology brought about by the ongoing discussion of the scientific status of multiverse proposals. An epistemic shift would involve changing the standard criteria for when a theory or model can be considered scientific – the most widespread such criterion being Popperian falsifiability. The main reason for claiming such an epistemic shift is the well-known problem that the multiverse may be untestable – or at least non-falsifiable – insofar as the other universes are in principle unobservable. Kragh places this problem in a historical context by showing a number of examples where cosmologists have argued for or against changing the rules of science in connection with new theoretical ideas. In the present day context, all cosmologists agree that testability is an indispensable and central criterion for cosmology as a science. Nevertheless, Kragh provides an informative discussion of how many different notions of testability are actually in play in the cosmological literature. As a consequence, Kragh shows that cosmologists are divided on the question of how (and at what stage in theory building) testability should be applied, and even what this criterion exactly means.

Jeremy Butterfield is concerned in his paper with various aspects of underdetermination in cosmology. He argues that cosmology in this regard is different from other sciences in that possible data in cosmology is limited to those of the Earth's past light-cone (as opposed to observable events anywhere in spacetime), and that underdetermination in cosmology is often about models, e.g. solutions of general relativity, rather than of underlying theories. This situation implies that the underdetermination of cosmological models is endemic: Butterfield discusses theorems which affirm that in most general relativistic spacetimes, we simply cannot gather sufficient data to observationally determine whether we live in one spacetime or another. Breaking this underdetermination requires making assumptions about the whole of spacetime, for instance by asserting the validity of the cosmological principle. However, Butterfield reviews the possible reasons for expecting this principle to hold outside our past-light cone and finds them lacking. Moreover, even the establishment of the cosmological principle within our past light-cone (the observable universe) is to some extent debatable. Butterfield concentrates his discussion on the more established parts of the cosmological standard model, from about 1 second after the Big Bang (the onset time of nucleosynthesis) to now. In the last section, he takes up the possible justification for the cosmological principle coming from the very early (inflationary) universe. As Butterfield hints, however, at best one may explain a feature of the late universe by invoking speculative physics (e.g. the conjectural inflaton field) at very early times.

The observational situation in present-day cosmology is addressed from three very different perspectives in the papers by Hamilton, López-Corredoira and Falkenburg. Jean-Christophe Hamilton gives a detailed account of the observational basis and status of the hot Big Bang model – in its currently favored version known as the Lambda Cold Dark Matter model – which is consistent with the widespread sense of correctness of the Big Bang picture. Yet, as Hamilton points out, triumphalism on behalf of the Big Bang is not in order. For there are still dark clouds on the horizon, most notably the well-known questions concerning the unknown origin and nature of dark matter and dark energy. Both of these elements, which are crucial to the matter-energy budget being consistent with a wide range of observations (in particular the WMAP results), are embarrassing from a theoretical point of view. Moreover, Hamilton points to some less well-known problems threatening the consistency between observations and Lambda-CDM predictions: The Lithium-7 problem which concerns a too low observed abundance of this isotope with respect to nucleosynthesis predictions, problems in accounting for some galactic scale dynamics, as well as possible anomalies in the Cosmic Microwave Background radiation. Nevertheless, in Hamilton's view, the overall agreement between the Lambda-CDM model and a wide range of increasingly accurate observations is very impressive. Thus all of the mentioned problems should be seen as important motivations for further investigations of the model rather than signs of defeat.

Martín López-Corredoira takes a rather more skeptical view on the observational situation. In his contribution, he first gives an informative overview of alternative cosmological models, including the quasi-steady state theory, plasma cosmology, static (or eternal) models, and a number of variations on the standard Big Bang model. To the extent these models are even considered by the majority of cosmologists, they are usually taken to be observationally ruled out. However, given that the Big Bang model also has its problems – and López-Corredoira gives a long list of such problems – things might not be that easy. In López-Corredoira's view, a plausible explanation for why the Big Bang model currently does better than its alternatives is simply that these alternatives are less developed than the former. This highly controversial claim – which amounts to a radical form of underdetermination in cosmology – rests on the idea that *ad hoc* additions may be invoked to make a model fit the cosmological data in case of conflict. The following sections contain a personal account of possible sociological factors – including groupthink and censorship in the electronic archives – which might affect the consensus opinion on the superiority of the Big Bang model. While López-Corredoira insists that he is not defending a social constructivist perspective and that there are objective data in cosmology, his paper is unmistakably polemical, and one referee felt that the paper was overly biased against standard cosmology. Nevertheless, I think the paper is important since it expresses a sentiment which is not often heard, and which should not be forgotten when discussing the philosophical and foundational aspects of modern cosmology.

In her paper, Brigitte Falkenburg takes up yet a different aspect of the observational situation in cosmology, namely the role of astroparticle physics. This field broadens the empirical basis of cosmology by taking into account also the information coming from cosmic rays – often analyzed in underground laboratories by particle physics methods. In Falkenburg's view, it is still not obvious precisely in what way this rather new discipline contributes to cosmology. In any case, she argues, astroparticle physics is philosophically interesting in that it challenges standard philosophy of science views e.g. of scientific explanation and realism. In particular, astroparticle physics pursues a 'bottom-up' approach in which models of the cosmic sources are constructed from data, and the study of subatomic particles are hoped to be informative about the large-scale structures of the universe. This contrasts with the 'top-down' approach of cosmology, more in line with standard views on explanation in philosophy of science, in which one proceeds from theory to data and from ideas of the large scale structures of the universe to their small scale observable consequences. Falkenburg points out that astroparticle physics pursues different heuristic unifying strategies – consistent with the fact that no unified theory of sub-atomic particles and their cosmic sources is known. This leads to an interesting combination of causal realism about the cosmic sources together with instrumentalism, or temporary ignorance, concerning the specific causal mechanisms giving rise to the observed spectrum of cosmic rays.

The last two papers concentrate on inflation and its consequences for early-universe and multiverse cosmology. Robert Brandenberger argues in his paper that inflation is not the well-established theory that it is often taken to be. Both because inflation still has several unresolved conceptual problems and because viable alternatives for theories of the early universe exist. The latter issue is yet another instance of underdetermination in modern cosmology. Brandenberger presents two scenarios – matter bounce and string gas cosmology – which, like inflation, may account for the origin of the structure in the universe, as observed in terms of inhomogeneities in the distribution of galaxies and small amplitude anisotropies in the cosmic microwave background radiation. Regarding the conceptual problems for inflation, Brandenberger pays special attention to the *trans-Planckian problem* for cosmological perturbations. The problem is that the prediction from inflation of the origin of structure depends on there being physical length scales smaller than the Planck length at the onset time of inflation, and so this prediction may well be sensitive to currently unjustified assumptions regarding unknown (quantum gravity) physics. This problem does not arise in the alternative models presented, although Brandenberger points out that these models have conceptual problems of their own. One will therefore have to make some sort of trade-off regarding strength and weaknesses of inflation and its competitors but, in any case, Brandenberger is confident that observations will be able to tell them apart in due time.

While Brandenberger questions the case for inflation by presenting alternatives, Chris Smeenk addresses the issue of how much weight can be put on the shoulders of inflationary cosmology, assuming that inflation indeed gives a roughly correct picture of the early universe. In particular, Smeenk asks whether inflationary cosmologists are entitled to draw far reaching conclusions concerning the existence of a multiverse. It transpires from Smeenk's discussion that typical arguments in favor of the multiverse, based on anthropic considerations, are hard to sustain, as they are based on a number of questionable 'ifs': Only if we could somehow count the elements of an infinite ensemble of pocket universes (the measure problem), and if we could then justify the introduction of probabilities, and if anthropic predictions could discriminate between competing theories, then one might come up with a prediction lending support to the multiverse. There is a remaining 'if', touched upon in the end of the article, which is the assumption that inflation actually leads to a multiverse. In fact, anthropic considerations regarding the multiverse have been invoked as a possible response to the fine-tuning problem for inflation (which holds that inflation needs more finely tuned parameters for the inflationary potential, than those of the fine-tuning problems inflation was originally meant to solve). The trouble is, as Smeenk points out, that this strategy may well undermine the motivations which led people to inflation in the first place, as one might use anthropic considerations to explain the cosmological fine-tuning (flatness and horizon) problems directly without inflation. In this case, the above mentioned problems of making anthropic predictions in support of the multiverse reappear as problems of making anthropic predictions in support of adequate initial conditions for our universe.

Before closing this introduction, let me come back to the above mentioned currently rising interest in the philosophical aspects of cosmology. Several possible reasons may be given for this. For instance, widely studied and open questions in philosophy of physics such as the nature of space-time, the possible unification in science, and the ramifications of various proposals for a quantum theory of gravity all find common ground within the philosophy of cosmology (see e.g. Ellis 2007). Moreover, cosmology is one of few scientific disciplines in which scientists openly and currently discuss philosophical issues, for example concerning the very nature of science itself. In addition, one could point to a reason which is not particularly new, although much exploited in contemporary popular writings on science: There seems to be something intrinsically fascinating about the cosmological questions, and they command a considerable public interest. In my view, this relates to another worthwhile philosophical aspect of cosmology – namely aesthetics – which is barely treated in the papers included here (see however Ellis' section 8).

Aesthetic considerations are by no means foreign to cosmology. 'Kosmos' in Greek may refer to order, adornment – as in 'cosmetics' – and also an orderly universe. As we know from Pythagoras, cosmology in this sense started out as the science of the harmonious and the beautiful; the science of a finite and well-ordered cosmos. Later on, and not least in the work of Giordano Bruno, the idea of an infinite universe entered the cosmological scene. In 1757, British philosopher Edmund Burke pointed out that the infinite is a main source for the aesthetic category known as the sublime.[2] And so, cosmology could be seen also as the science of the sublime; the science of uncountable island universes extending towards infinity. No wonder, then, that cosmology has always had such a strong aesthetic appeal. And, if we are to believe Burke and the power of the infinite, no wonder that the multiverse idea is so attractive to many modern day cosmologists. On this point, Bersanelli (2011) makes an interesting observation. An infinite universe is not automatically aesthetically attractive. If the spatially infinite is merely an indefinite repetition of what is already known, it might well stop to surprise

---

[2] "The ideas of eternity and infinity are among the most affecting we have; and yet perhaps there is nothing of which we really understand so little, as of infinity and eternity." (Burke 1757, section 4)

us. And even a multiverse in which everything that can happen always happens may be boring in the end: "Nothing really happens in a world where everything always goes on infinitely often… It seems that spatial infinity, in order to be perceived as a fascinating concept, has to maintain some kind of element of selected variety and genuine surprise" (Bersanelli 2011, 205).

In any case, as also testified by the papers in this volume, there are no reasons to become disillusioned with modern cosmology. We do not know, and probably cannot ever come to know, whether the universe is in fact infinite. More generally, given principled limitations on possible data, if not those stemming from our own finitude, cosmological knowledge will always be incomplete. Thus cosmology and its philosophical study can safely be expected to keep offering surprises.

**Acknowledgements**
I would like to thank the speakers and contributors to this volume for their participation and patience, general editor of this journal Dennis Dieks for his help and encouragement, and my colleague Svend E. Rugh for support and discussions along the way. Thanks also to the Spanish Ministry of Science and Innovation (Project FFI2011-29834-C03-02) for financial support.

**References**
- Bersanelli, M. (2011). "Infinity and the Nostalgia of the Stars". In M. Heller y W. Hugh Woodin (eds.) *Infinity: New Research Frontiers*. Cambridge: Cambridge University Press, 193-217.
- Burke, E. (1757): *A Philosophical Enquiry into the Origin of Our Ideas of the Sublime and Beautiful.* London: Routledge and Paul, 1958.
- Ellis, G. F. R. (2007). "Issues in the philosophy of cosmology." In *Handbook for the Philosophy of Physics,* J. Earman and J. Butterfield (eds.), Part B. Amsterdam: Elsevier, 1183-1286.
- Rugh, S. E. and Zinkernagel, H. (2011). "Weyl's principle, cosmic time and quantum fundamentalism". In D. Dieks *et al*. (eds.), *Explanation, Prediction and Confirmation*. The Philosophy of Science in a European Perspective, Berlin: Springer Verlag, 411-424.
- Smeenk, C. (2013). "Philosophy of cosmology". In R. Batterman (ed.), *Oxford Handbook of Philosophy of Physics*. New York, NY: Oxford University Press, 607-652.